\documentclass[lettersize,journal]{IEEEtran}

\renewcommand{\arraystretch}{0.95}

\IEEEoverridecommandlockouts
\usepackage{booktabs}
\usepackage{siunitx}
\usepackage{amsfonts}
\usepackage[dvips]{graphicx}
\usepackage{times}
\usepackage{cite}
\usepackage{amsmath}
\usepackage{cases}
\usepackage{array}
\usepackage{dsfont}
\usepackage{amssymb}

\usepackage{makecell}
\usepackage{stfloats}
\usepackage{booktabs}
\usepackage{graphicx}
\usepackage{subfigure}
\usepackage{pdfpages}
\usepackage{footnote}
\usepackage{soul}
\usepackage{array}
\usepackage{multirow}
\usepackage{bm}
\usepackage{empheq}
\usepackage{amsthm}
\usepackage{algorithm}
\usepackage{algorithmicx}
\usepackage{algpseudocode}
\usepackage{color}
\usepackage{diagbox}
\usepackage{caption}
\theoremstyle{plain}

\theoremstyle{definition}

\usepackage[colorlinks,bookmarksopen,bookmarksnumbered,citecolor=blue, linkcolor=blue, urlcolor=blue]{hyperref}

\let\oldReturn\Return
\renewcommand{\Return}{\State\oldReturn}

\usepackage{textcomp}
\usepackage{xcolor}
\begin{document}
	\title{Channel-Aware Preemptive Scheduling for Semantic Communication with Truncated Diffusion and Path Compensation}
	\vspace{-4pt}
	
	\vspace{-25pt}\author{\IEEEauthorblockN{$\text{Chengyang Liang}$, $\text{Dong Li}, ~\IEEEmembership{Senior Member,~IEEE}$}
		\vspace{-22pt}
		\thanks{Chengyang Liang and Dong Li are with the School of Computer Science and Engineering, Macau University of Science and Technology, Macau, China (e-mail: 3240006992@student.must.edu.mo; dli@must.edu.mo).}}
	\vspace{-20pt}

\maketitle

\begin{abstract}
	Semantic communication (SemCom) presents a transformative paradigm for alleviating bandwidth limitations in mobile networks by transmitting task-relevant semantic features instead of raw data bits. While SemCom systems utilizing diffusion models achieve superior generation quality, existing research treats semantic generation and wireless transmission as temporally independent processes. This separation neglects the intrinsic conflict between the multi-step iterative delays inherent in diffusion models and the time-varying fading characteristics of wireless channels. To address this discrepancy, this paper proposes a channel-aware preemptive scheduling with truncated diffusion and path compensation (CAPS-TDPC) framework. Contrary to conventional methods that require completion of the generation phase prior to transmission, the proposed framework implements a channel-driven scheduling mechanism: each user maintains a countdown inversely proportional to its instantaneous channel gain, and the user with the shortest countdown transmits immediately, regardless of whether its diffusion process has completed. This design permits the interruption of the forward diffusion process to enable “early” transmission under favorable channel conditions. In addition, a receiver-side compensation mechanism grounded in path dynamics is developed to mitigate the semantic loss resulting from such interruptions. A path deficit metric is proposed at the receiver to quantify the recovery difficulty of distinct image blocks by incorporating the velocity field of the inverse dynamics model, which allows for adaptive weighted inverse sampling. Experimental evaluations demonstrate that the proposed framework substantially reduces the end-to-end latency while maintaining the high-fidelity semantic reconstruction, thereby enhancing the system robustness in fast fading channel environments.
\end{abstract}

\begin{IEEEkeywords}Semantic communication, channel-aware preemptive scheduling, diffusion models, truncated diffusion, path compensation.
\end{IEEEkeywords}

\IEEEpeerreviewmaketitle

\vspace{-4pt}
\section{Introduction} 
\label{sec:Introduction}

\IEEEPARstart{T}{he} rapid expansion of intelligent applications, including autonomous driving, extended reality (XR), and the metaverse, presents unprecedented challenges for wireless networks in terms of data throughput and latency reduction \cite{GU, LessData, FromData, GenAIsemcomLiang, Onlinesemcom}. Conventional communication frameworks, based on Shannon's information theory, emphasize bit-level accuracy in data transmission but often neglect the semantic significance and task relevance of the transmitted content \cite{Asurveysemcom, ResourceAllocation}. In contrast, within the context of next generation networks, characterized by human-machine interaction and intelligent decision-making, achieving bit-level precision is both inefficient and unnecessary. Semantic Communication (SemCom) represents a paradigm shift by employing deep learning (DL) techniques to extract and transmit semantically meaningful features efficiently \cite{JSCNA, SemanticSuccessive, ChannelCalibration}. This approach has achieved high performance for image, speech, and text transmission, particularly under conditions of low signal-to-noise ratio (SNR) \cite{Aunified, NLJSCNSC}.

In recent years, advancements in generative artificial intelligence (GenAI) have  accelerated the development of SemCom. Diffusion Probabilistic Models (DPMs) have become a pivotal technology for semantic codec design, owing to their robust distributional modeling capabilities and superior generative performance \cite{Semanticimportance, SCcarrier}. In our prior work, we investigated the active incorporation of channel noise into the diffusion process, thereby demonstrating the integration of wireless foundation for generative models. Nonetheless, current diffusion-based SemCom systems encounter a fundamental temporal limitation: diffusion models (DMs) necessitate dozens or even hundreds of iterative denoising steps to produce high-quality images \cite{LatentDiffSemCom, Alightweight, ImageGeneration}. In mobile environments, this extended generation latency inherently conflicts with the millisecond-scale channel coherence time \cite{OptimizingResource, CommunicateLess, LightweightDM}.

Existing multi-user (MU) semantic resource allocation methods derive scheduling metrics solely from channel state information (CSI) or static semantic importance, treating the generative model as a black box \cite{Semanticimportance, MUSemantic, OptimizingResource}. For example, Liang et al. proposed semantic‑importance‑aware mapping over multi-input multi-output (MIMO) fading channels \cite{Semanticimportance}, while Liu et al. optimized multi‑modal semantic transmission using a diffusion‑based game approach \cite{CommunicateLess}. Lang et al. further introduced a meaning‑first scheduling framework that jointly optimizes bandwidth, power, and edge computing resources under eavesdropping threats \cite{Semanticsecurity}. However, these methods lack awareness of the internal diffusion state and therefore cannot determine whether early transmission is feasible when channel conditions become favorable before the diffusion process completes \cite{MUGenSemcom, SemanticPriorAided, AdaptiveAwareness}. In edge computing scenarios, computational and communication resources are inherently coupled, yet most studies optimize them independently. This decoupling limits semantic fidelity under constrained latency and computational budgets \cite{WirelessVision, JointMixed, SemComOPTIMA}. Recent works have attempted to reduce diffusion steps via latent diffusion models \cite{LatentDiffSemCom}, lightweight‑to‑diffusion switching \cite{Alightweight}, or model quantization \cite{LightweightDM}.

Although generative model-based SemCom systems have achieved notable advancements in improving reconstruction quality, they still encounter a fundamental trade-off between the generation latency and the time-varying characteristics of wireless channels in practical mobile deployments. Most existing diffusion-based SemCom systems adopt a sequential paradigm, in which they complete the entire forward diffusion process before transmission. This static pipeline overlooks CSI, leading to a critical dilemma: while a user waits for the diffusion process to finish, a previously favorable channel condition may deteriorate, causing transmission failures or degraded reconstruction quality. To address this gap, this paper proposes a channel-aware preemptive scheduling with truncated diffusion and path compensation (CAPS-TDPC) framework. Specifically, at the transmitter, each user maintains a countdown inversely proportional to its instantaneous channel gain. The user whose countdown reaches zero first is selected to transmit immediately, regardless of whether its diffusion process has completed, sending the current intermediate semantic features. At the receiver, to recover from these incomplete features caused by early transmission, we introduce a path compensation mechanism based on flow matching, which solves an ordinary differential equation to evolve the features backward from the intermediate state to the full semantic distribution. Furthermore, to optimize reconstruction under limited computational budgets, we design a patch-wise adaptive reconstruction allocation strategy that dynamically assigns reverse steps based on each patch’s semantic importance and recovery difficulty. This strategy significantly enhances semantic reconstruction while maintaining limited computational overheads.

The contributions of this paper are summarized as follows.

\begin{itemize}
	\item \textbf{Channel-Aware Preemptive Scheduling Framework:} To address the temporal mismatch between diffusion generation delay and channel coherence time, this paper introduces a channel-aware preemptive scheduling framework that enables early transmission. Specifically, each user maintains a countdown inversely proportional to its instantaneous channel gain. The user whose countdown reaches zero first transmits its intermediate semantic features immediately, regardless of whether the diffusion process has completed. This mechanism prioritizes channel quality over generation completion, transforming the conventional generation-then-transmission paradigm into a channel-driven opportunistic transmission. As a result, it effectively reduces the end-to-end semantic transmission delay while avoiding the time cost of waiting for the whole diffusion process to finish under deteriorating channel conditions.
	\item \textbf{Path Dynamics-Based Compensation:} To address incomplete semantic evolution caused by premature transmission, we introduce a flow-matching-based path compensation framework. The generative process lost due to diffusion interruption is characterized as a path deficit on the latent semantic manifold, quantified by the velocity field of the inverse dynamics model. This deficit serves as a metric for semantic recovery difficulty. We then formulate the receiver-side semantic reconstruction as an initial value problem of ordinary differential equations constrained by channel conditions, enabling continuous trajectory completion from any intermediate state. The proposed method overcomes the limitation of traditional DMs that require a purely stochastic initial state, ensuring semantic consistency and recoverability under incomplete transmission.
	\item \textbf{Patch-wise Adaptive Reconstruction:} Given the significant heterogeneity in semantic importance and restoration difficulty across different regions of an image, this study proposes a fine-grained, patch-level adaptive inverse diffusion mechanism. This mechanism integrates semantic attention weights with patch loss metrics to formulate a ratio metric representing semantic gain relative to computational cost. This metric is then employed to dynamically allocate the number of inverse diffusion steps and computational resources, thereby enabling semantic value-driven diffusion scheduling. In contrast to conventional uniform inversion strategies, the proposed method prioritizes the restoration of regions exhibiting high semantic value and restoration efficiency within a constrained computational budget. As a result, it substantially enhances the reconstruction quality of critical semantic regions under conditions of low SNR and severe fading, while achieving an improved performance-complexity trade-off.
\end{itemize}

The remainder of this paper is structured as follows. Section \ref{sec:system model} introduces the system model and formulates the problem. Section \ref{sec:architecture} describes the proposed system, including the scheduling strategy, path dynamics modeling, and patch-wise adaptive resource allocation. Section \ref{sec:Experiments} presents numerical experiments and ablation studies to assess the effectiveness of the proposed framework. Finally, Section \ref{sec:conclusion} concludes the paper.

\vspace{-4pt}
\section{System Model}
\label{sec:system model}
\vspace{-2pt}

\begin{figure*}[htp]
	\vspace{-8pt}
	\centering
	\includegraphics[scale=0.77]{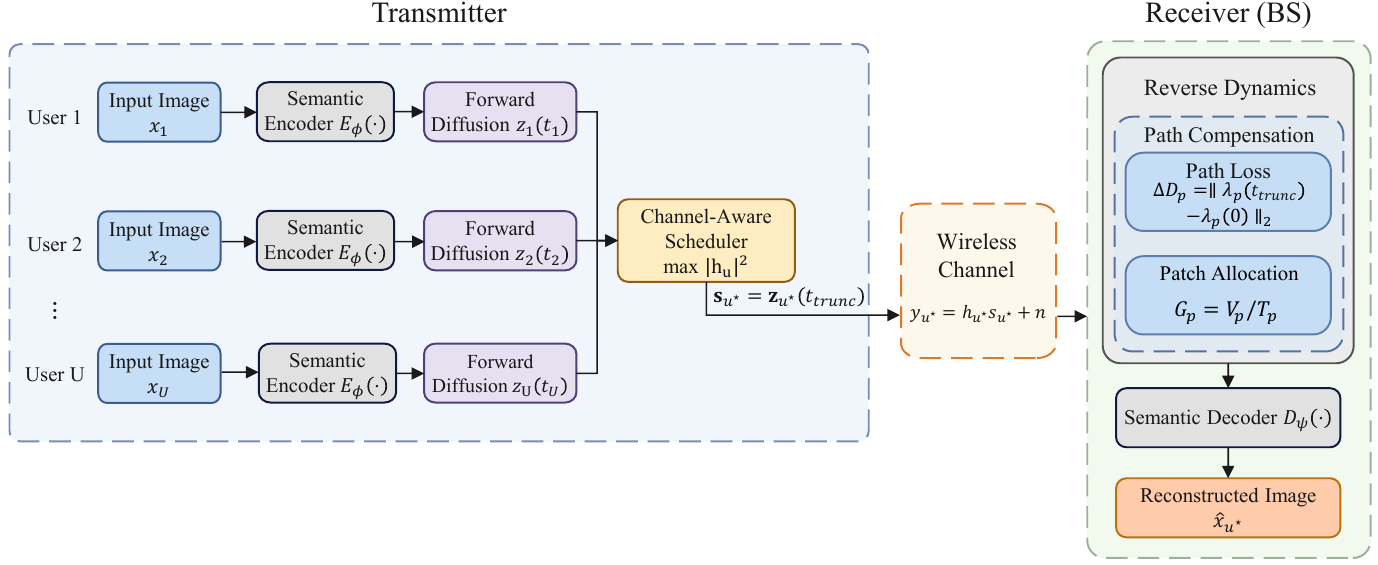}
	\caption{Illustration of the proposed channel-aware preemptive scheduling with truncated diffusion and path compensation semantic communication system.}
	\label{fig:system}
	\vspace{-10pt}
\end{figure*}

This section presents a comprehensive system model for the proposed CAPS-TDPC SemCom framework as illustrated in Fig \ref{fig:system}. We consider an MU uplink SemCom system, which comprises three primary components: transmitter semantic modeling coupled with forward diffusion transmitter, channel-aware preemptive scheduling integrated with channel knowledge, and receiver path compensation alongside the semantic reconstruction. By accounting for the temporal dimensions of both the generative process and the physical communication channel, the system facilitates early transmission prior to the completion of diffusion and ensures the accurate restoration of complete semantic information at the receiver.

\vspace{-10pt}
\subsection{Transmitter Model and Forward Diffusion}
\label{sec:transmitter}

Consider the system containing a set of $U$ users, indexed as $u \in \mathcal{U} = \{1, 2, \dots, U\}$. In a given transmission task, each user $u$ transmits a high-dimensional red, green, blue (RGB) image to the base station (BS), denoted as $\mathbf{x}_u \in \mathbb{R}^{H \times W \times 3}$.

\subsubsection{Semantic Encoding}
Before undergoing the diffusion process, the image $\mathbf{x}_u$ is initially processed by a semantic encoder $\mathcal{E}_\phi(\cdot)$ based on a visual transformer (ViT). This encoder compresses the high-dimensional pixel space image into a compact latent semantic representation

\begin{equation}
	\mathbf{z}_u = \mathcal{E}_\phi(\mathbf{x}_u) \in \mathbb{R}^{P \times d},
\end{equation}

\noindent where $P$ denotes the number of feature patches into which the image is divided, and $d$ represents the semantic dimension of each feature patch. 

\subsubsection{Forward Diffusion}
Unlike traditional source coding, this system considers the semantic feature $\mathbf{z}_u$ as the initial state $\mathbf{z}_u(0)$ of the DM. The discrete set of diffusion time steps is defined as $t \in \{0, 1, \dots, T\}$. The forward diffusion process gradually perturbs the latent representation by injecting Gaussian noise according to a predefined variance schedule. $\beta_{t}\in(0,1)$ denotes the variance schedule at timestep $t$, and define $\alpha_{t}=1-\beta_{t}$. The cumulative product is defined as $\bar{\alpha}_t=\prod_{s=1}^t\alpha_s$. The diffusion process is then expressed as

\begin{equation}
	\mathbf{z}_u(t) = \sqrt{\bar{\alpha}_t}\, \mathbf{z}_u(0) + \sqrt{1 - \bar{\alpha}_t}\, \epsilon,
\end{equation}

\noindent where $\epsilon \sim \mathcal{N}(0, 1)$ denotes the standard Gaussian diffusion noise, while $\alpha_t$ is a parameter defined by the predetermined noise scheduling variance. Throughout this procedure, all feature patches consistently share the identical forward diffusion time step $t_u$. We distinguish between two types of noise in the system:
(1) diffusion noise $\epsilon\sim\mathcal{N}(0,1)$, introduced in the generative process;
(2) channel noise $n\sim\mathcal{CN}(0,\sigma_{n}^{2})$, introduced for wireless transmission.

\subsection{Channel-Aware Preemptive Scheduling and Channel Transmission}
\label{sec:CAPS}

The physical layer wireless channel is characterized as a narrowband flat fading channel. During a given transmission time slot, the channel fading coefficient from user $u$ to the BS is $h_u \sim \mathcal{CN}(0, \sigma_h^2)$. To optimize the reconstruction gain and minimize the generation delay, the system initiates transmission prior to the completion of the diffusion process. Instead, we implement a channel-driven preemptive scheduling mechanism. The BS selects the user with the best instantaneous channel condition to transmit immediately.

The system assesses a channel clock for each user, defined as $T_u^c=1/|h_u|^2$, which reflects the instantaneous channel quality. A smaller $T_u^c$ indicates a better channel and a shorter countdown.

The BS determines the user for transmission by employing the scheduling metric $S_u=T_u^c=\frac{1}{|h_u|^2}$. Upon selection, the designated user promptly terminates the forward diffusion process and transmits the current intermediate state semantic feature as the transmission symbol $\mathbf{s}_u$, defined as $\mathbf{s}_u = \mathbf{z}_u(t_{trunc})$. $t_{trunc}$ represents the truncated time step at the moment of scheduling. The received signal at the BS is expressed as follows

\begin{equation}
y_u = h_u s_u + n,
\end{equation}

\noindent where $n \sim \mathcal{CN}(0, \sigma_n^2)$ denotes the complex additive white Gaussian noise (AWGN).

\subsection{Receiver Path Compensation and Semantic Reconstruction}
\label{sec:Receiver}

\subsubsection{Dynamic Path Compensation}
The receiver replaces conventional static decoders with a reverse dynamics model grounded in flow matching to compensate for the semantic loss and recover the complete semantic information from an interrupted intermediate state. We distinguish between two time variables in the system: the discrete diffusion timestep $t \in \{0,1,\dots,T\}$ used in the forward process, and the continuous time variable $\tau$ used to model the reverse dynamics via ordinary differential equations (ODEs). The parameterized inverse dynamics model is denoted as $v_\theta(\mathbf{s}, \tau, h)$, which produces the velocity field on the feature manifold conditioned on the current state, time index, and channel state. The receiver then evolves the features backward by solving the following dynamic equation

\begin{equation}
	\frac{d\mathbf{s}}{d\tau} = v_{\theta}(\mathbf{s},\tau ,h).
\end{equation}

This continuous dynamic process enables the BS to compensate for semantic distortion caused by early transmission. Operationally, this equation is iteratively adapted at the patch level via a discrete sampler. The specific patch-level adaptive compensation algorithm, which is based on path loss metrics, will be described in the following section.

\subsubsection{Image Reconstruction}

Upon completion of the inverse dynamics compensation process, when reaching $t=0$, the system acquires the recovered pure latent semantic feature set $\hat{\mathbf{z}} = \{\mathbf{z}_p(0)\}_{p=1}^P$. Subsequently, this feature set is input into a pre-trained semantic decoder $\mathcal{D}_\psi(\cdot)$ to reconstruct the target image, $\hat{\mathbf{x}}_u = \mathcal{D}_\psi(\hat{\mathbf{z}})$, which can be utilized for downstream tasks at the receiver for human visual perception.

Through channel-aware preemptive scheduling with truncated diffusion process at the transmitter and dynamic path compensation at the receiver, the system forms a complete closed-loop, achieving a coordinated cross-layer mapping between physical-layer transmission resources and semantic-layer generation steps.

\vspace{-6pt}
\section{Channel-Aware Preemptive Scheduling with Truncated Diffusion and Path Compensation Mechanism}
\label{sec:architecture}

In the proposed CAPS-TDPC framework, the transmitter truncates the forward diffusion process immediately upon user scheduling. Although this early termination reduces end-to-end latency, it inevitably results in incomplete semantic features. Consequently, the received signal contains only partially diffused latent representations mixed with channel noise. Applying conventional reverse diffusion from such an intermediate, non-stationary state results in severe reconstruction degradation. To recover high-fidelity semantic information without sacrificing low latency, we propose a receiver-side path compensation mechanism.

Flow matching \cite{lipman2022flow} is a continuous-time generative framework that transports a prior distribution to the target data via an ordinary differential equation. However, standard flow matching assumes the reverse process starts from a known prior, which cannot be directly applied to  our truncated diffusion scenario, in which the receiver obtains an intermediate, non-stationary state corrupted by incomplete diffusion and channel noise. To overcome this, we adapt flow matching to enable reverse evolution from any intermediate point by training a conditional inverse dynamics model. This allows us to reconstruct high-fidelity semantic features from a prematurely interrupted diffusion. The following subsections elaborate on the scheduling logic, the compensation algorithm, and the adaptive resource allocation.

\subsection{Channel-Aware Preemptive Scheduling with Truncated Diffusion}
\label{sec:channel-Clock}

In the system model, we have established the fundamental framework of the scheduling metric. This section explores the dynamic execution logic of this mechanism within MU systems. Traditional SemCom systems assume that the execution of the generation model is independent of the physical channel state, resulting in a substantial mismatch between the generation latency and the channel coherence time. To address this, we propose a preemptive scheduling mechanism based on CSI.

We reformulate the BS scheduling problem as a channel-driven priority rule. The user with the best instantaneous channel condition is selected for immediate transmission. The BS executes the dynamic scheduling policy, selecting the user $u^\star$ with the minimum penalty metric to allocate the current time. $t_u$ denote the current diffusion step of user $u$ at the scheduling moment. Once user $u^\star$ is selected, we set $t_{trunc} = t_{u^\star}$, the user immediately stops its forward diffusion at its current step.

\begin{equation}
	u^\star = \arg\min_{u \in \mathcal{U}} \frac{1}{|h_u|^2}.
\end{equation}

\noindent The primary advantage of this scheduling logic lies in its preemptive nature. When a user's channel condition rapidly improves to an optimal state $T_{u^\star}^c \to 0$, that user is selected immediately, even if its diffusion process has just started. Conversely, when all users experience poor channel conditions, the system may wait for better channel opportunities; the diffusion progress is not considered in scheduling decisions. This design avoids the risk of waiting for diffusion completion under deteriorating channel conditions.

This approach involves applying a truncation operator to the forward diffusion Markov chain associated with the user. We define the instantaneous truncated diffusion step at the scheduling time as $t_{trunc}$. The user immediately terminates all subsequent forward noisy propagation computations from $t_{trunc}$ to $T$ and extracts the current intermediate latent variable as the baseband transmission symbol $\mathbf{s}_{u^\star}$

\begin{equation}
	\mathbf{s}_{u^\star} = \Gamma\big(\mathbf{z}_{u^\star}(t)\big)\Big|_{t=t_{trunc}} = \mathbf{z}_{u^\star}(t_{trunc}).
\end{equation}

\noindent The system substantially decreases the generation delay for users experiencing high channel quality by employing the truncation operator. Nevertheless, the transmission advantage afforded by the proposed scheduling mechanism incurs a compromise in the integrity of the transmitter features. The signal received is affected not only by physical channel noise $\mathbf{n}$ but also represents a non-stationary intermediate distribution that deviates from the pure semantic distribution of the intended target. Therefore, a pivotal theoretical challenge for this framework is to accurately compensate for the prematurely truncated evolutionary trajectory at the receiver.

\subsection{Path Loss Dynamics Modeling Based on Flow Matching Models}
\label{sec:path loss}

As previously indicated, the proposed scheduler initiates early transmission at time $t_{trunc}$. The received equalized signal $\hat{\mathbf{s}}_{u}$ at the BS constitutes an intermediate semantic feature that is corrupted by the physical channel and has not undergone complete diffusion. Conventional Markov-based discrete diffusion denoising models require the reverse process to commence strictly from pure Gaussian noise ($t=T$) with a fixed step size \cite{DiffusionSC}. These models are inadequate for recovery tasks that start at arbitrary intermediate time $t_{trunc}$ and exhibit non-stationary distribution characteristics. 

\begin{figure}[tp]
	\vspace{-2pt}
	\centering
	\includegraphics[scale=0.45]{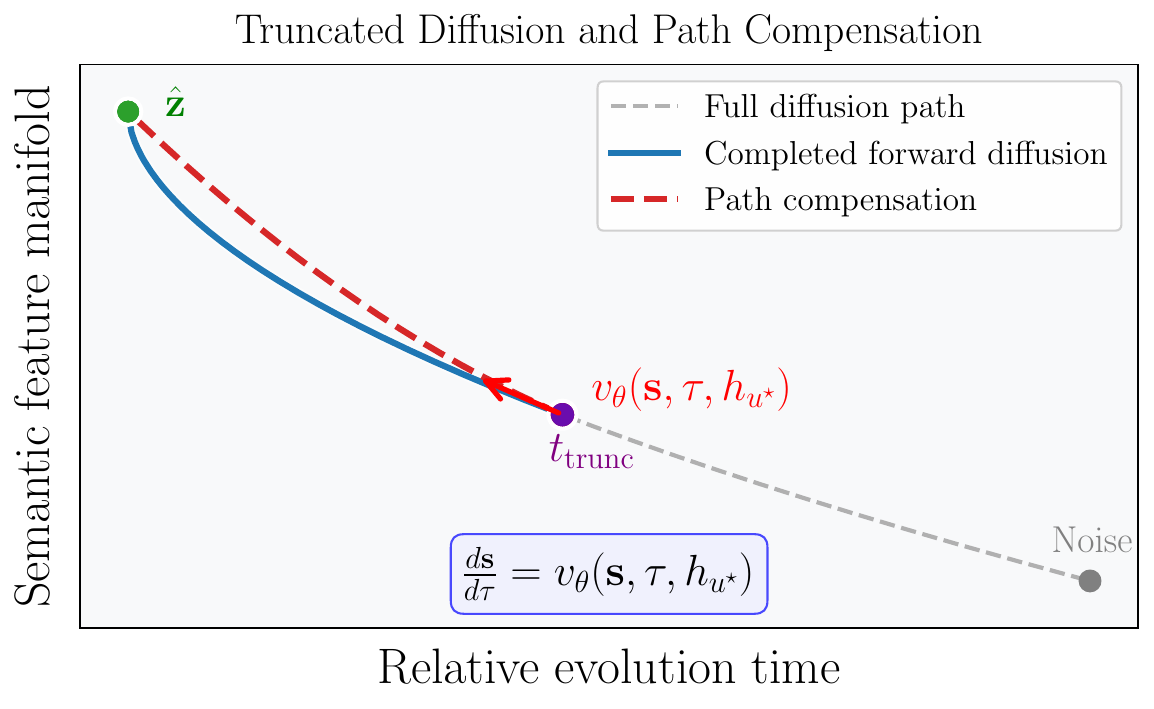}
	\caption{Illustration of the proposed path compensation mechanism.}
	\label{fig:path_compensation}
	\vspace{-4pt}
\end{figure}

As illustrated in Fig.~\ref{fig:path_compensation}, although all feature patches within the same image experience identical transmission interruption times $t_{trunc}$, the semantic distances that each patch must traverse to return to the pure data manifold differ substantially. This disparity arises from pronounced semantic density imbalances within the image, such as the contrast between complex textured regions and smooth background areas. Allocating uniform backpropagation resources across all segments results in overfitting in background regions, which generates artifacts, and underfitting in detailed regions, leading to blurred features. Therefore, we need a mechanism that can not only compensate for the incomplete trajectory but also adapt to the heterogeneous recovery difficulty across patches.

To address these challenges, we depart from the traditional discrete denoising framework and instead formulate the semantic reconstruction process at the receiver as an initial value problem for ODEs in the continuous-time domain. Let the continuous time variable be $\tau \in [0, t_{trunc}]$. We interpret feature recovery at the receiver as a deterministic diffeomorphic evolution of feature points on a manifold within the latent space. To model this evolution, we introduce a parameterized inverse dynamics model $v_\theta(\mathbf{s}, \tau, h)$ to fit the velocity field along the evolution path. The core function of this network is to predict the tangent direction of the feature manifold at a given time step. For the selected user $u^\star$, the compensation path of the received feature $\hat{\mathbf{s}}_{u^\star}$ is described by the following ordinary differential equation

\begin{equation}
	\frac{d\mathbf{s}}{d\tau} = v_\theta(\mathbf{s}, \tau, h_{u^\star}),
\end{equation}

where the input to the dynamic model $v_\theta$ includes not only the current feature state $\mathbf{s}$ and the time index $\tau$, but also explicitly incorporates the physical channel state $h_{u^\star}$ as a conditioning variable. The continuous time variable $\tau$ is defined over the interval $[0, t_{trunc}]$, where $t_{trunc}$ is the discrete diffusion step at which transmission occurs. The physical rationale for incorporating the channel state $h_{u^\star}$ is that channels exhibiting different levels of fading cause signal distortions with varying variances. The dynamic model must be capable of recognizing the channel SNR associated with the current feature in order to produce an appropriately corrected and denoised velocity output.

To quantify the varying recovery difficulty across different patches, we propose a novel metric termed the path deficit. For the $p$th feature chunk $\mathbf{s}_p$, its path deficit $\Delta D_p$ is defined as the $\ell_2$ norm integral of the inverse dynamics velocity field over the remaining unscheduled time interval $[0, t_{trunc}]$

\begin{equation}
	\Delta D_p = \int_{0}^{t_{trunc}} \big\| v_\theta\big(\mathbf{s}_p(\tau), \tau, h_{u^\star}\big) \big\|_2 \, d\tau.
\end{equation}

Here $\tau$ denotes continuous time corresponding to the reverse evolution from the truncated state at $t_{trunc}$ back to $t=0$. From the perspective of geometric topology, the integral term $\int \|v_\theta\| d\tau$ precisely measures the path length along which the $p$th feature chunk transitions from the intermediate truncated state $\mathbf{s}_p(t_{trunc})$ back to the true semantic distribution $\mathbf{s}_p(0)$ along the optimal manifold path. A greater path length signifies a higher degree of missing semantic information in that region, thereby increasing the difficulty of restoration. Accordingly, $\Delta D_p$ serves as a rigorous mathematical metric to quantify the semantic penalty incurred by each chunk as a result of premature transmission.

Since continuous-time ODEs cannot be solved directly on digital processors, numerical solvers are employed at the receiver to approximate discretization. The inverse solution process is partitioned into $K$ discrete steps, where the current discrete step index is $k \in \{1, 2, \dots, K\}$. The corresponding discrete timestamp is $t_k$, and the integration step size is $\eta_k = t_{k-1} - t_k$. The discrete form of the path deficit for the $p$th feature block can then be computed as

\begin{equation}
	\Delta D_p \approx \sum_{k=1}^{K} \eta_k \big\| v_\theta(\mathbf{s}_p^k, t_k, h_{u^\star}) \big\|_2.
\end{equation}

The receiving BS concurrently computes the $\ell_2$ norm of each patch feature vector and accumulates these values over the temporal dimension following each reverse step velocity prediction, thereby dynamically generating the deficit value for each patch. This approach departs from the uniform calculation and allocation mechanisms characteristic of traditional DMs by incorporating a path loss model. The introduction of $\Delta D_p$ enables the system not only to determine the necessity of transmission but also to precisely identify the locations where remedial actions are most required at the receiving end. This development provides a robust theoretical foundation for the design of a path deficit-based adaptive reconstruction algorithm, which is presented in the subsequent section.

\subsection{Semantic-Latency Guided Patch-wise Reconstruction}
\label{sec:Adaptive Patches Algorithm}

Following the initiation of early transmission via the proposed scheduling mechanism, the receiver obtains hybrid features $\hat{\mathbf{s}}_{u^\star}$ truncated at the diffusion intermediate step $t_{trunc}$ and subject to channel corruption. The principal challenge confronting the receiver at this juncture is to reconstruct high-quality images from these degraded features under finite computational constraints $\mathcal{B}_{\mathrm{max}}$, such as the maximum permissible number of backward passes or decoding latency. Traditional DMs impose a uniform number of backward iterations across all image blocks. This static strategy exhibits fatal flaws in mobile computing scenarios. On one hand, challenging regions containing complex high-frequency textures consume substantial computational resources yet may still underfit. On the other hand, smooth background areas converge extremely rapidly, yet contribute minimally to overall semantic understanding.

To optimize the allocation of computational resources at the receiver, this section introduces an image-block adaptive reverse scheduling mechanism based on joint semantic-latency gain. This mechanism extends the scheduling concept, originally applied at the transmitter, to the receiver in a symmetrical manner. By utilizing a combined metric that balances the semantic importance of each image block against its decoding time cost, the mechanism effectively prioritizes the allocation of computational resources to specific blocks.

For each image patch $p\in\{1,2,\ldots,P\}$ within the receiver's latent feature map $\hat{\mathbf{z}}\in\mathbb{R}^{P\times d}$, we define two key metrics. One is the semantic value $\mathcal{V}_{p}$, which quantifies the importance of image patch $p$ to the overall semantic understanding of the image. This metric is directly derived from the self-attention weights obtained from the transmitter-side semantic encoder $\mathcal{E}_{\phi}$ as the basis for this metric. Let $a_{u,p}$ denote the attention weight assigned to block $p$ by user $\text{u}$'s encoder, subject to the global normalization condition $\sum_{p=1}^Pa_{u,p}=1$. We then define $\mathcal{V}_p=a_{u,p}. $ Higher attention weights generally correspond to foreground objects, text, or semantically salient regions within the image, which are critical for achieving high-quality reconstruction.

Another metric considered is the decoding time cost, denoted as $\mathcal{T}_p=\Delta D_p$. This metric estimates the computational overhead necessary to restore block $p$ from its current corrupted state to a clean data manifold. Based on the path loss modeling described in Section III-B, we directly utilize the path deficit $\Delta D_{p}$ as the cost metric.

\noindent Path deficit quantifies the path length of the path required for block $p$ to return from its truncated state $\mathbf{s}_p(t_{trunc})$ to its original semantic distribution $\mathbf{s}_p(0)$ on the latent space manifold. According to flow matching theory, this path length is strongly positively correlated with the number of integration steps needed in the inverse process. Therefore, a larger value of $\mathcal{T}_{p}$ signifies increased difficulty in recovering the block and results in longer decoding times. These two metrics characterize the intrinsic properties of image blocks: semantic importance and computational overhead. Their combined consideration provides a principled basis for diffusion scheduling at the receiver.

To facilitate effective scheduling decisions within limited budgetary constraints, it is essential to employ a scalar metric that quantitatively captures the semantic benefit per unit of computational resource. Drawing inspiration from the transmitter's channel-aware preemptive scheduling, we define the scheduling benefit $\mathcal{G}_p$ as the ratio of semantic value to decoding time cost: $\mathcal{G}_p = \frac{\mathcal{V}_p}{\mathcal{T}_p}$. This metric encapsulates the principle of balancing semantic significance against decoding latency efficiency. Blocks exhibiting high semantic value coupled with low decoding time yield the highest gain and should be prioritized for processing. Conversely, blocks characterized by low semantic value or substantial decoding costs, resulting in low gain, can be deferred or omitted under stringent budgetary limitations.


From an optimization perspective, the optimal solution to the problem of maximizing the total semantic value under budget constraints, in its continuous relaxed form, entails allocating resources in descending order according to $\mathcal{V}_p/\mathcal{T}_p$. In our discrete framework, although selection based on ratio sorting does not necessarily ensure global optimality, it is computationally efficient and provides a practical approximation of the optimum. This characteristic renders it particularly suitable for the real-time scheduling requirements of edge devices. More importantly, this formulation exhibits perfect symmetry with the transmitter scheduler. The transmitter selects users by minimizing the $T_{u^\star}^c$, while the receiver selects blocks by maximizing $\mathcal{V}_p/\mathcal{T}_p$ – both approaches reflect a unified prioritization strategy based on joint metrics.

Given a computational budget $\mathcal{B}_{\max}$, the receiver first obtains $\nu_{p}$ for each block $p$ based on pre-stored attention weights, and calculates $\mathcal{T}_{p}$ using the path loss model. Subsequently, the scheduling benefit $\mathcal{G}_p=\mathcal{V}_p/\mathcal{T}_p$ is computed. Next, all blocks are sorted in descending order of $\mathcal{G}_{p}$ to form a priority list $\mathcal{L}$. The block with the highest gain is placed at the top, indicating its optimal cost-effectiveness. The remaining budget is initialised as $\mathcal{B}_{\mathrm{rem}}=\mathcal{B}_{\mathrm{max}}$, and the selected block set is initialised as $\mathcal{S}=\emptyset$. Each block $p$ in the list $\mathcal{L}$ is traversed sequentially. If $\mathcal{T}_p\leq\mathcal{B}_{\mathrm{rem}}$, allocate the full number of reverse steps to this block and add $p$ to $\mathcal{S}$, deducting $\mathcal{T}_{p}$ from the remaining budget $\mathcal{B}_{\mathrm{rem}}=\mathcal{B}_{\mathrm{rem}}-\mathcal{T}_p$. Otherwise, cease the allocation.

\subsection{Training Strategy}
\label{sec:Training Strategy}

In the practical implementation of the CAPS-TDPC framework, the proposed scheduler dynamically determines the truncation time $t_{trunc}$ based on the instantaneous channel. Concurrently, the receiver utilizes the proposed semantic-latency-driven, patch-wise adaptive reconstruction scheduling to allocate the computational budget $\mathcal{B}_{\mathrm{max}}$. To ensure effective operation under this scheduling policy, we have developed an end-to-end training strategy that simulates the complete transmitter-receiver pipeline, incorporating the adaptive receiver-side scheduling.

The training of traditional DMs is generally conducted on clean data manifolds using fixed noise schedules, with the objective of minimizing Gaussian prediction errors at each step. Nevertheless, this framework faces challenges due to the presence of highly coupled, incomplete diffusion noise alongside the residual physical channel noise, which together produce compounded distortions at the receiver. To mitigate these issues, we incorporate transmission interruption behavior and channel fading as random variables during the training process. For each training image $x$ in a mini-batch, we execute the following simulation.

First, we sample a transmission interruption time $t_{trunc}\sim\mathcal{U}(0,T)$ from a discrete uniform distribution, forcing the model to learn from incomplete diffusion paths of arbitrary lengths. Second, given the predefined physical channel model $\mathcal{H}$, the instantaneous channel coefficient $h$ is randomly sampled according to $\mathcal{H}$. Third, we sample a computational budget $\mathcal{B}_{\max}$ uniformly sampled from a feasible range $[B_{\min},B_{\max}]$ to mimic varying resource constraints at the edge. Forward diffusion is applied to the latent representation $\mathbf{z}(0)=\mathcal{E}_\phi(\mathbf{x})$ up to the step $t_{trunc}$, obtaining $\mathbf{z}(t_{trunc})$.

Then simulate channel transmission and equalization to obtain the received corrupted feature $\hat{\mathbf{s}}=\begin{pmatrix}h\cdot\mathbf{z}(t_{trunc})+n\end{pmatrix}/h$, where $n\sim\mathcal{CN}(0,\sigma_n^2)$ is the channel noise. We compute the norm of the velocity field predicted by the current inverse dynamics model $v_{\theta}$ at the initial state and approximate the total path deficit as $\hat{\mathcal{T}}_p=\eta\|v_\theta(\hat{\mathbf{s}}_p,t_{trunc},h)\|_2$, where $n$ is a scaling factor proportional to $t_{trunc}$. This provides a rough estimate of the required computational effort. Using the semantic scores $\mathcal{V}_p=a_{u,p}$ obtained from the encoder’s attention, compute the scheduling benefit $\mathcal{G}_p=\mathcal{V}_p^\alpha/\hat{\mathcal{T}}_p^\beta$ for each patch. Sort patches by descending $\mathcal{G}_{p}$ and select the top patches whose total estimated cost does not exceed the budget $\mathcal{B}_{\mathrm{max}}$, forming the selected set $\mathrm{S}$. Perform the full reverse ODE integration for the patches in $\mathrm{S}$ only, using a numerical solver for $K$ steps. Patches not in $\mathrm{S}$ remain at their current corrupted state $\hat{\mathbf{s}}_{p}$. The reverse process yields the recovered latent features $\hat{\mathbf{z}}^{(0)}$. Pass $\hat{\mathbf{z}}^{(0)}$
through the semantic decoder $\mathcal{D}_{\psi}$ to obtain the reconstructed image 
$\hat{x}$. The overall training objective is to minimize the expected pixel-level reconstruction error $\mathcal{L}(\phi,\theta,\psi)=\mathbb{E}_{\mathbf{x},t_{trunc},h,\mathcal{B}_{\mathrm{max}}}\left[\left\|\mathcal{D}_\psi(\hat{\mathbf{z}}^{(0)})-x\right\|_2^2\right].$

This comprehensive simulation enables the network to adapt its inverse dynamics predictions across diverse truncation times, channel conditions, and computational budgets, while the attention mechanism and path deficit estimation co-evolve to facilitate effective gain-based scheduling. The training procedure is outlined in Algorithm \ref{alg:CAPS-TDPC}.

\begin{algorithm}[t]
	\caption{End-to-End Training Algorithm for CAPS-TDPC Framework with Gain-Based Scheduling}
	\label{alg:CAPS-TDPC}
	\begin{algorithmic}[1]
		\Require $\mathcal{X}, T, K, \mathcal{H}, [B_{\min},B_{\max}], \alpha,\beta$
		\Ensure Trained $\phi,\theta,\psi$
		\State Initialize $\phi,\theta,\psi$
		\While{not converged}
		\State Sample $\mathbf{x}\sim\mathcal{X}$; $\mathbf{z}_0,\mathbf{a}=\mathcal{E}_\phi(\mathbf{x})$
		\State Sample $t\sim\mathcal{U}(0,T)$, $h\sim\mathcal{H}$, $B_{\max}\sim\mathcal{U}(B_{\min},B_{\max})$
		\State $\mathbf{z}_t=\sqrt{\bar{\alpha}_t}\mathbf{z}_0+\sqrt{1-\bar{\alpha}_t}\epsilon$, $\hat{\mathbf{s}}=(h\mathbf{z}_t+\mathbf{n})/h$
		\State For each $p$: $\hat{\mathcal{T}}_p=\frac{t}{K}\|v_\theta(\hat{\mathbf{s}}_p,t,h)\|_2$, $\mathcal{G}_p=\mathbf{a}_p^{\alpha}/\hat{\mathcal{T}}_p^{\beta}$
		\State Sort patches by $\mathcal{G}_p$ descending; $\mathcal{S}=\emptyset$, $B_{\text{rem}}=B_{\max}$
		\For{$p$ in sorted list}
		\If{$\hat{\mathcal{T}}_p\le B_{\text{rem}}$} $\mathcal{S}\cup\{p\}$, $B_{\text{rem}}= \hat{\mathcal{T}}_p$
		\Else \textbf{break}
		\EndIf
		\EndFor
		\State $\mathbf{s}=\hat{\mathbf{s}}$
		\For{$k=1$ to $K$}
		\State $t_k,\eta_k\leftarrow$ current; $\mathbf{v}=v_\theta(\mathbf{s},t_k,h)$
		\For{$p\in\mathcal{S}$} $\mathbf{s}_p=\eta_k\mathbf{v}_p$ \EndFor
		\EndFor
		\State $\hat{\mathbf{x}}=\mathcal{D}_\psi(\mathbf{s})$; $\mathcal{L}=\|\hat{\mathbf{x}}-\mathbf{x}\|_2^2$
		\State Update $\phi,\theta,\psi$ via backprop
		\EndWhile
		\State \Return $\phi,\theta,\psi$
	\end{algorithmic}
	\end{algorithm}

\section{Numerical Experiments}
\label{sec:Experiments}
In this section, we evaluate the proposed CAPS-TDPC framework through comprehensive numerical experiments conducted under diverse channel conditions and MU scenarios. The objective is to demonstrate that the proposed system can effectively balances transmission latency and reconstruction quality by employing channel-aware preemptive scheduling and patch-wise adaptive reconstruction, thereby outperforming conventional SemCom approaches. We first introduce the datasets and baseline methods, followed by a description of the experimental configurations. Subsequently, we present quantitative comparisons, visual analyses, and ablation studies to elucidate the contributions of the key components within the CAPS-TDPC framework.

\vspace{-8pt}
\subsection{Experiments Setting}
\subsubsection{Dataset}

The proposed CAPS-TDPC framework is evaluated using two benchmark image datasets of differing resolutions to examine its adaptability across varying levels of complexity. The CIFAR-100 dataset consists of 60,000 natural images, each measuring $32 \times 32$ pixels, spanning 100 distinct object categories. The ImageNet-256 dataset utilizes a subset of the large-scale ImageNet dataset, with images resized to $256 \times 256$ pixels. This higher-resolution context poses considerable challenges for semantic transmission, necessitating that the DM effectively capture intricate spatial structures and texture patterns while maintaining an efficient latent representation.

For all datasets, images are normalized to the range $[-1, 1]$ via the transformation $(x - 0.5) / 0.5$ to conform to the input specifications of the DM. The integration of these datasets facilitates a comprehensive evaluation of CAPS-TDPC across both low-resolution and high-resolution visual domains, thereby substantiating the generalizability of the proposed scheduling and adaptive allocation mechanisms.

\subsubsection{Parameters Setting}
All experiments are implemented in PyTorch 2.1.0 with CUDA 12.1 on a Linux server equipped with an NVIDIA A100-80GB GPU. The CAPS-TDPC model consists of a ViT-based semantic encoder $E_\phi$, a forward diffusion module, a reverse dynamics model $v_\theta$ based on flow matching, and a convolutional decoder $D_\psi$.

The semantic encoder employs a ViT architecture with patch size 16, embedding dimension $d = 256$, and depth 6, which extracts compact latent representations from input images. For CIFAR-100, the encoder produces $P = 4$ patches with $d = 256$ channels each. For ImageNet-256, it generates $P = 256$ patches. The reverse dynamics model $v_\theta$ consists of 8 transformer blocks with cross-attention layers conditioned on channel state $h$ and timestep $t$. The decoder employs a U-Net architecture with adaptive upsampling blocks, automatically adjusting the number of upsampling stages based on input resolution, 2 stages for $32 \times 32$ and 4 stages for $256 \times 256$.

The forward diffusion process follows a linear noise schedule with $T = 1000$ total timesteps during training. However, leveraging the proposed scheduling, inference adaptively terminates diffusion at timestep $t_{trunc}$ based on instantaneous channel quality, significantly reducing the transmission latency. We employ the AdamW optimizer with an initial learning rate of $1 \times 10^{-4}$ and cosine annealing scheduling over 1000 epochs. Mixed precision training is enabled to accelerate computation and reduce memory consumption. The batch size is resolution-adaptive: 128 for CIFAR-100 and 16 for ImageNet-256, ensuring stable gradient updates while maximizing GPU utilization. Gradient clipping with a unit norm is applied to prevent training instability. The loss function integrates pixel-level $\ell_2$ reconstruction error with perceptual loss weighted by $\lambda_{perceptual} = 0.1$ to enhance visual quality.

Channel coefficients are modeled as Rayleigh fading $h \sim \mathcal{CN}(0, \sigma_h^2)$ with varying SNR levels $\{0, 5, 10, 15, 20, 25\}$ dB to simulate diverse wireless conditions. The receiver employs patch-wise adaptive reconstruction sampling guided by the scheduling benefit $G_p = V_p / T_p$, where semantic value $V_p$ is derived from encoder attention weights and decoding cost $T_p$ is computed via path deficit integration $\Delta D_p = \int_0^{t_{trunc}} \|v_\theta(s_p(\tau), \tau, h)\|_2 d\tau$. High-gain patches receive proportionally more reverse steps within a maximum budget of 200 steps per patch, enabling efficient reconstruction prioritized toward semantically critical and easily recoverable regions.

Reconstruction quality is evaluated using the Peak SNR (PSNR), Multi-Scale Structural Similarity Index (MS-SSIM), and Learned Perceptual Image Patch Similarity (LPIPS), which respectively quantify pixel-level accuracy, structural fidelity, and perceptual quality. Transmission latency is measured in diffusion timesteps, representing the computational and temporal overhead prior to channel transmission. All metrics are averaged over 100 independent trials to ensure statistical robustness, with error bars indicating the standard deviation across runs.

\subsubsection{Baseline Schemes}
To comprehensively validate the proposed CAPS-TDPC framework, we compare against representative baselines spanning traditional separation-based coding, deep learning-based joint source-channel coding, and recent diffusion-based SemCom methods. All baselines operate under identical multi-user channel conditions and TDMA structures to ensure fair comparison.

\begin{itemize}
	\item \textbf{Traditional JSCC (JPEG + LDPC)}, where each user independently applies JPEG compression followed by LDPC channel coding. The BS sequentially decodes each user's transmission without exploiting semantic features. This baseline represents the conventional separation-based digital communication pipeline and serves as a performance lower bound.
	
	\item \textbf{DeepJSCC} \cite{DeepJSCC}, an end-to-end trainable joint source-channel coding scheme where convolutional encoders directly map images to channel symbols, and decoders reconstruct images from noisy signals. Unlike CAPS-TDPC, this baseline does not employ DMs or adaptive scheduling, representing state-of-the-art non-generative deep learning approaches.
	
	\item \textbf{Fixed-Timestep Diffusion (Fixed t=500)}, a diffusion-based system where all users execute forward diffusion to a fixed timestep $t_{fixed} = 500$ before transmission, regardless of channel quality. The receiver applies uniform reverse denoising without patch-wise adaptation. This baseline isolates the contribution of scheduling by removing adaptive transmission timing.
	
	\item \textbf{CDDM} (Channel Denoising Diffusion Models) \cite{CDDM}, a diffusion-based physical layer module is positioned subsequent to channel equalization to model the distribution of the channel input signal and to mitigate residual channel noise. When integrated with JSCC, CDDM serves as a representative of diffusion-based SemCom systems that do not incorporate scheduling or patch-wise adaptive reconstruction.
	
	\item \textbf{WITT} (Wireless Image Transmission Transformer) \cite{WITT}, a transformer-based joint source-channel coding scheme utilizing Swin Transformers as the backbone to effectively capture long-range dependencies in high-resolution image transmission. A spatial modulation module adjusts latent representations based on CSI, allowing a single model to accommodate varying channel conditions. In contrast to CAPS-TDPC, the WITT framework does not integrate DMs or adaptive scheduling mechanisms.
\end{itemize}

The proposed CAPS-TDPC fundamentally differs from all baselines through three key mechanisms: (1) \textit{channel-aware preemptive scheduling} dynamically determines transmission timing based on the channel countdown $T_c^u$, enabling early transmission during favorable channel states; (2) \textit{path deficit quantification} via $\Delta D_p = \int_0^{t_{trunc}} \|v_\theta(s_p(\tau), \tau, h)\|_2 d\tau$ precisely measures semantic recovery difficulty, enabling accurate reconstruction from interrupted diffusion; (3) \textit{patch-wise adaptive allocation} balances semantic importance against decoding cost through $G_p = V_p / T_p$, achieving efficient quality-latency trade-offs.

\begin{figure}[tp]
	\vspace{-2pt}
	\centering
	\includegraphics[scale=0.202]{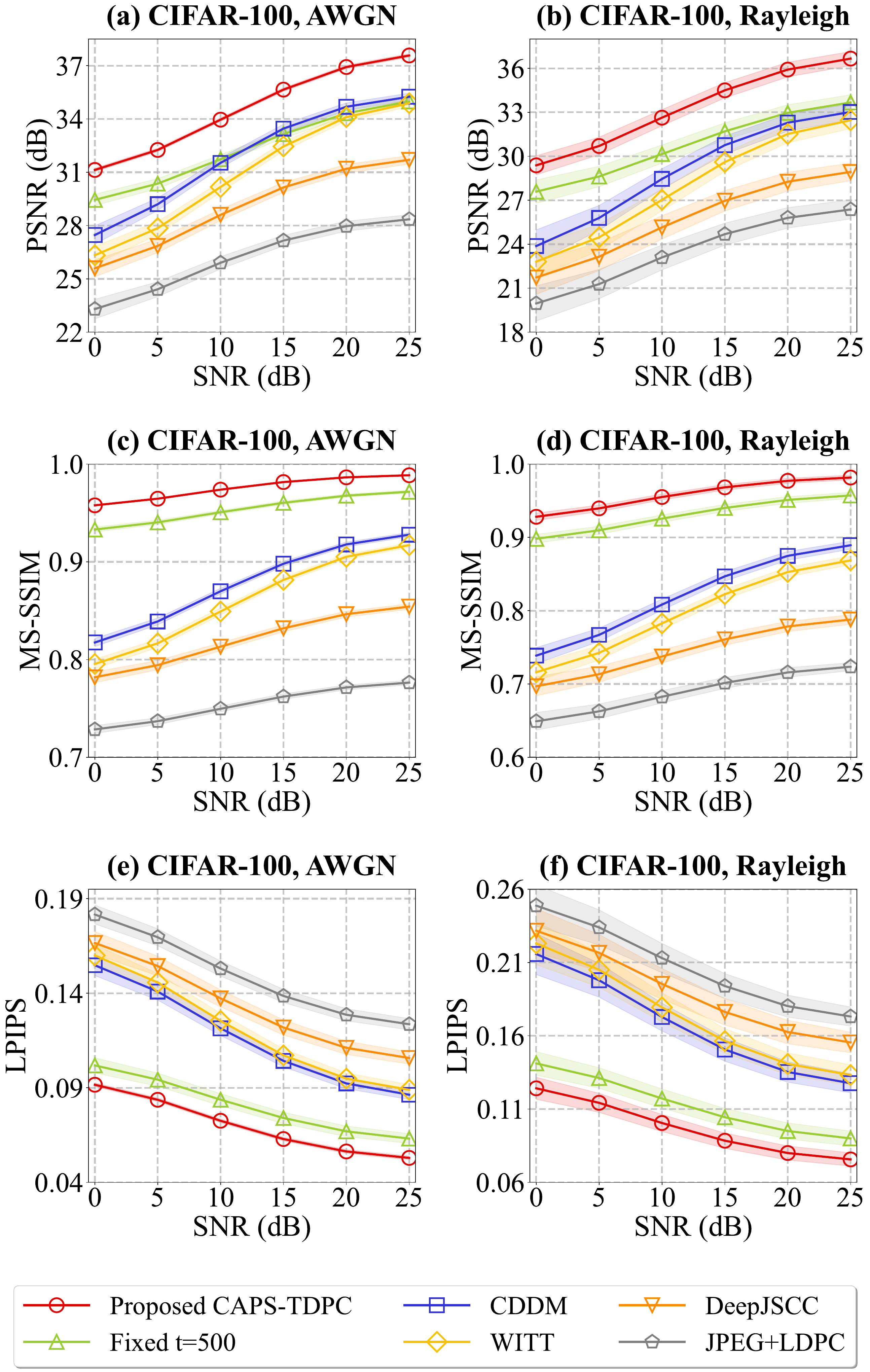}
	\caption{Comparison of the performance in Cifar-100 datasets with AWGN and Rayleigh fading.}
	\label{fig:cifar_comparison}
	\vspace{-4pt}
\end{figure}

\begin{figure}[tp]
	\vspace{-2pt}
	\centering
	\includegraphics[scale=0.202]{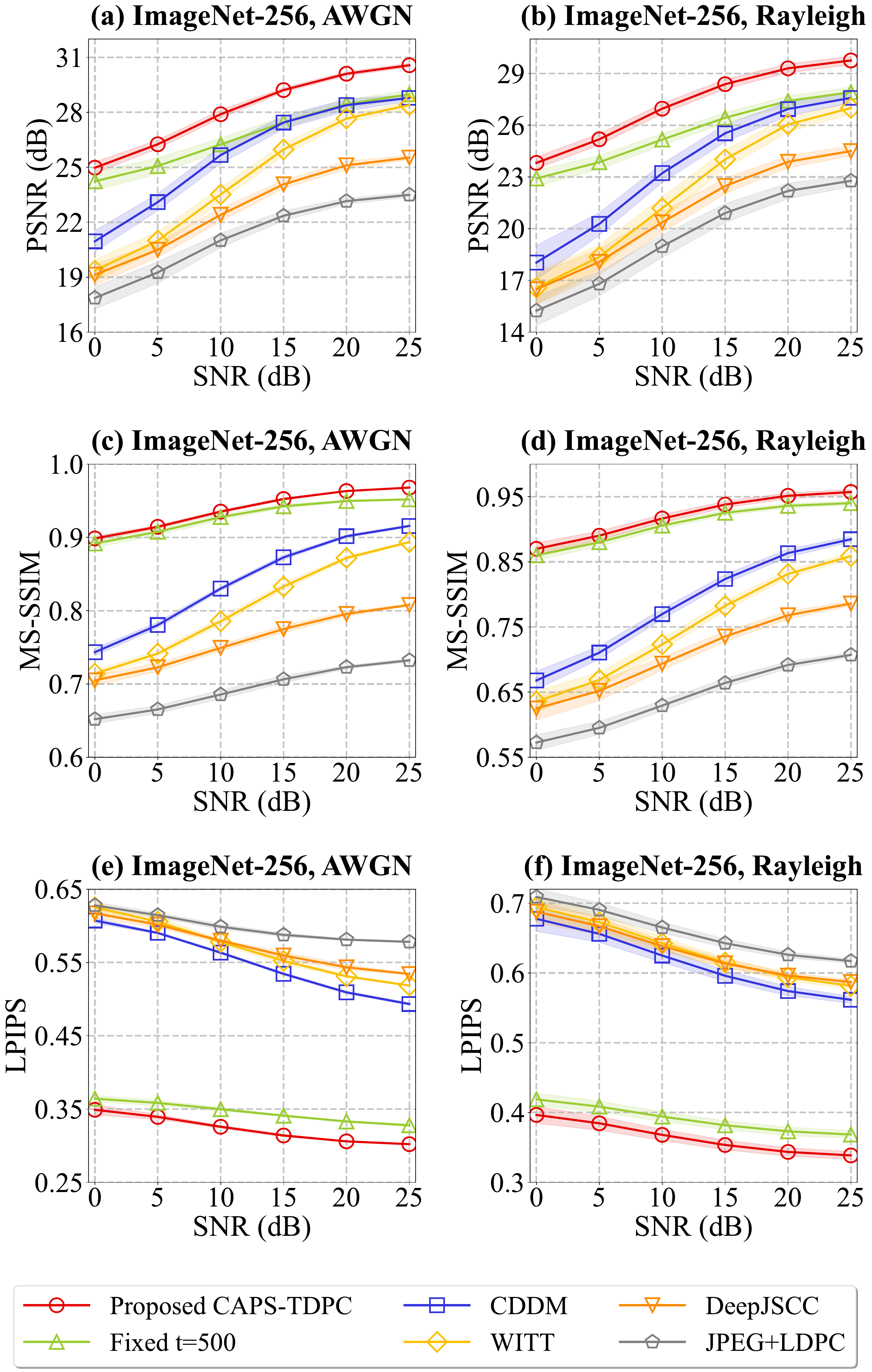}
	\caption{Comparison of the performance in ImageNet-256 with AWGN and Rayleigh fading.}
	\label{fig:imagenet_comparison}
	\vspace{-4pt}
\end{figure}

\vspace{-2pt}
\subsection{Performance Analysis}
\vspace{-2pt}

To comprehensively evaluate the effectiveness of the proposed CAPS-TDPC framework, this section presents a quantitative comparison between CAPS-TDPC and six representative baseline methods under AWGN and Rayleigh fading channel. The baseline methods include JPEG combined with LDPC, DeepJSCC, CDDM, WITT, and a fixed parameter setting of $t = 500$.

\subsubsection{PSNR Performance Analysis}

The top two rows of Figs. \ref{fig:cifar_comparison} and \ref{fig:imagenet_comparison} present the PSNR performance of each method on the CIFAR-100 and ImageNet-256 datasets, respectively. Overall, the PSNR values of all methods increase monotonically with rising SNR. However, CAPS-TDPC consistently demonstrates a clear advantage across all SNR levels and under both channel conditions. Under the AWGN channel, at low SNR values, CAPS-TDPC achieves an improvement of approximately 8 dB over traditional JSCC and about 5 dB over DeepJSCC on CIFAR-100. As the SNR increases beyond 20 dB, this advantage remains stable at approximately 2 to 3 dB. In the more challenging Rayleigh fading channel, due to random fluctuations in channel conditions, the absolute PSNR of all methods decreases; nevertheless, the performance gap between CAPS-TDPC and other SemCom methods further widens. At 15 dB, this gap exceeds 4 dB, indicating that CAPS-TDPC exhibits greater robustness to channel fading. It is noteworthy that diffusion-based baselines such as CDDM perform comparably to Fixed t=500 in the high SNR region but consistently underperform relative to CAPS-TDPC, thereby validating the effectiveness of the combined channel-aware preemptive scheduling and path compensation mechanisms.

\subsubsection{MS-SSIM and LPIPS Performance Analysis}

The results obtained from MS-SSIM and LPIPS metrics further substantiate the aforementioned conclusions. MS-SSIM, which assesses structural consistency, indicates that under AWGN conditions, the CAPS-TDPC method attains a score exceeding 0.98 in the high SNR range, whereas the traditional JSCC method fails to reach 0.80. Under the Rayleigh fading channel, the MS-SSIM scores of all evaluated methods generally decline by 0.05 to 0.10. However, CAPS-TDPC consistently maintains the highest values. Notably, at high SNR levels, its MS-SSIM remains close to 0.96, outperforming other baseline approaches. Regarding LPIPS, a perceptual distance metric where lower values indicate better performance, CAPS-TDPC achieves the lowest scores across both datasets and channel conditions. For instance, considering the ImageNet-256 dataset under a Rayleigh channel with 20 dB noise, CAPS-TDPC attains an LPIPS value of approximately 0.078, whereas the Fixed t=500 method exceed 0.09 and 0.10. This finding suggests that images restored by CAPS-TDPC are perceptually closer to the original images. Collectively, these results demonstrate that CAPS-TDPC not only improves the pixel-level accuracy but also consistently enhances the structural preservation and perceptual quality.

A comparison of the results obtained from CIFAR-100 and ImageNet-256 datasets indicates that the advantages of CAPS-TDPC are consistently maintained across both scenarios, with a more substantial improvement observed in high-resolution images. This enhancement can be attributed to its path-loss-based, block-wise adaptive allocation mechanism, which effectively manages complex textural regions. Moreover, although all methods exhibit reduced performance under the Rayleigh channel compared to the AWGN channel, CAPS-TDPC demonstrates the least performance degradation. Notably, the decline in the LPIPS metric is less severe than that observed in other baseline methods, underscoring the robustness of the proposed scheduling mechanism in fast-fading environments.

In summary, CAPS-TDPC consistently surpasses existing baseline methods across all evaluation metrics and experimental conditions. Its primary contributions lie in the synergistic integration of joint channel-aware preemptive scheduling, path loss compensation, and semantic-delay-driven block-by-block reconstruction, which collectively facilitate high-quality semantic reconstruction under low-latency transmission conditions. This provides an effective solution for SemCom in mobile scenarios.

\subsubsection{Patch-wise Reconstruction Analysis}

\begin{figure*}[htp]
	\vspace{-2pt}
	\centering
	\includegraphics[scale=0.47]{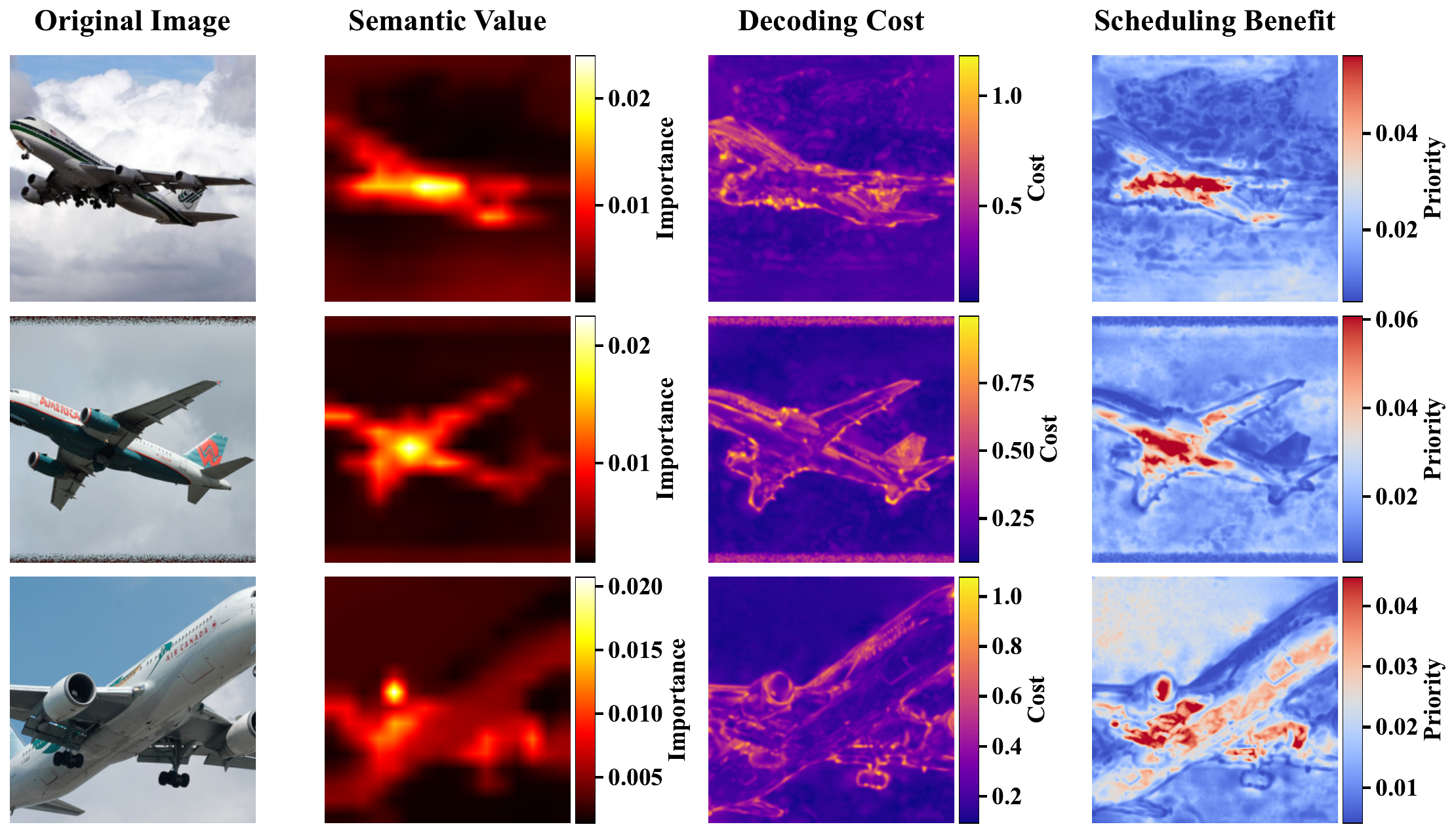}
	\caption{Illustration of the proposed semantic-latency guided patch-wise reconstruction.}
	\label{fig:heatmap}
	\vspace{-4pt}
\end{figure*}

To demonstrate the block-wise adaptive reconstruction mechanism within the proposed CAPS-TDPC framework, Fig. \ref{fig:heatmap} displays heatmaps of the semantic value $V_{p}$, decoding cost $T_p=\Delta D_p$, and scheduling benefit $G_p=V_p/T_p$ for three representative images. Each row presents, from left to right, the original image followed by the corresponding heatmaps of $V_{p}$, $T_{p}$, and $G_{p}$.

The decoding cost heatmap demonstrates considerable spatial heterogeneity. Regions characterized by complex textures exhibit high $T_{p}$, indicating increased recovery difficulty, whereas smooth backgrounds display low $T_{p}$. The coefficient of variation attains 16.6$\%$, underscoring the importance of adaptive reconstruction. The semantic value heatmap reveals that foreground objects, such as aircraft, receive higher attention weights. Notably, $V_{p}$ and $T_{p}$ exhibit a strong negative correlation, suggesting that semantically significant regions are easier to recover, which consists with the intuition that salient structures exhibit greater resilience to noise during diffusion. The scheduling benefit $G_{p}$ provides a unified priority metric. Regions with high gain correspond to blocks that are both semantically important and readily recoverable, whereas those with low gain are deprioritized. The dynamic range of $G_{p}$ reaches 0.06, facilitating fine-grained scheduling. Furthermore, a strong positive correlation between $G_{p}$ and $V_{p}$ confirms that semantic importance remains the predominant factor, while the negative $V_{p}-T_{p}$ correlation between $V_{p}$ and $T_{p}$ validates that prioritizing high-gain blocks optimizes semantic fidelity under constrained computational budgets.

In summary, the proposed semantic-latency-driven adaptive allocation mechanism effectively differentiates computational resources among image regions, as a result, an enhanced balance between transmission latency and reconstruction quality.

\vspace{-2pt}
\subsection{Ablation Studies}

\subsubsection{Ablation Studies on Scheduling}

\begin{figure}[tp]
	\vspace{-2pt}
	\centering
	\includegraphics[scale=0.43]{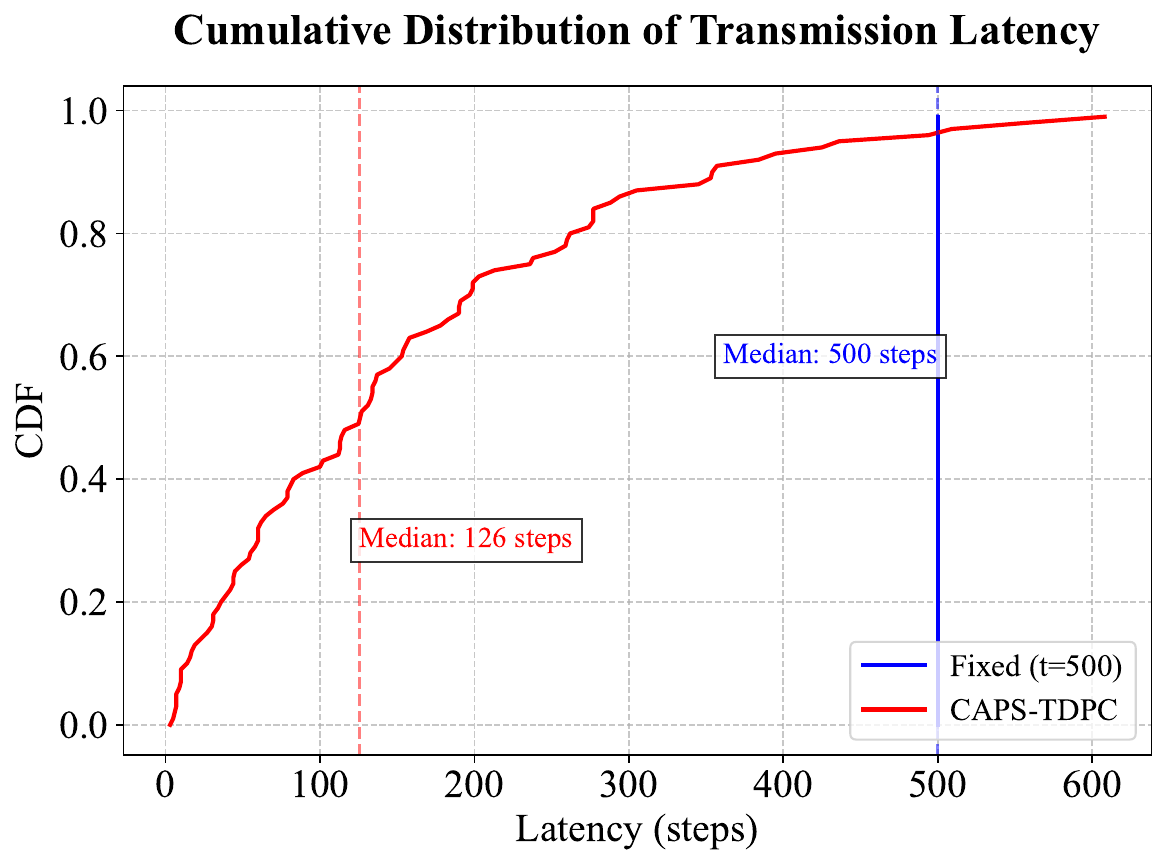}
	\caption{Cumulative distribution of transmission latency under Rayleigh fading.}
	\label{fig:Latency CDF}
	\vspace{-4pt}
\end{figure}

\begin{figure}[tp]
	\vspace{-2pt}
	\centering
	\includegraphics[scale=0.43]{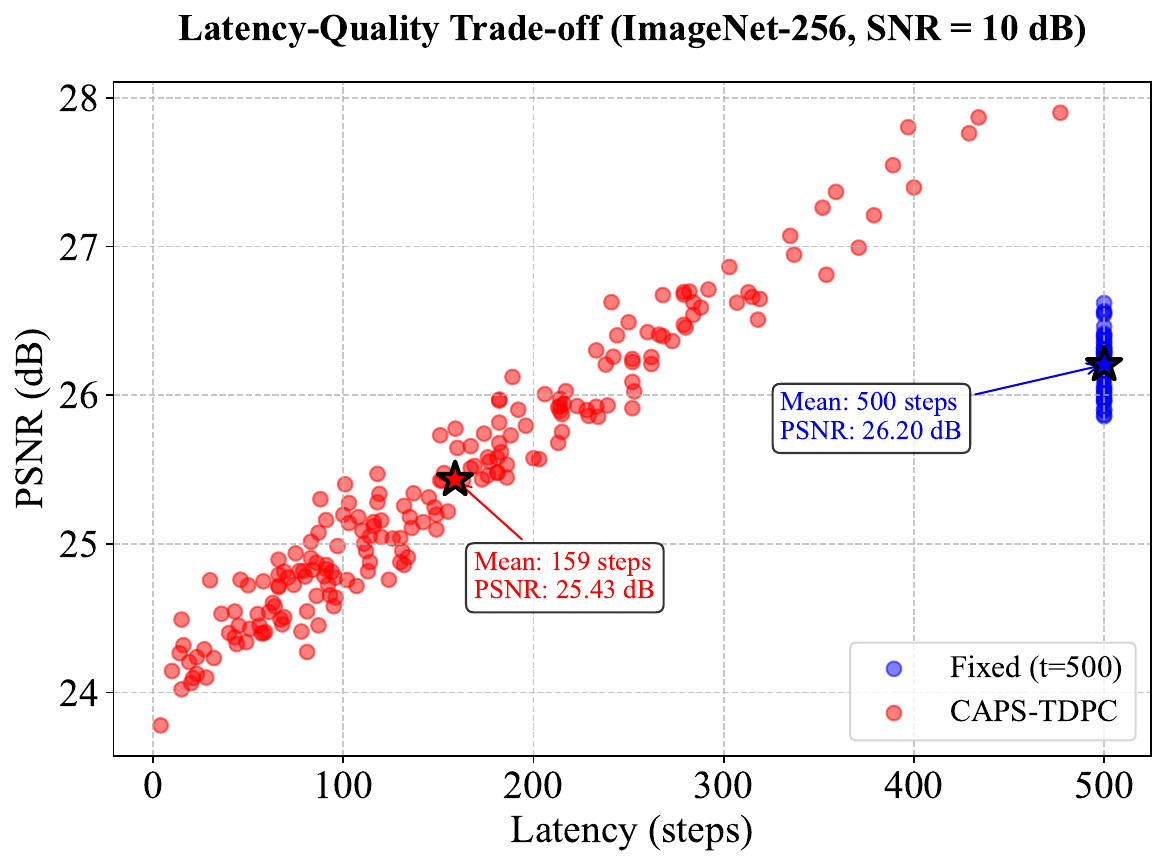}
	\caption{Latency-quality trade-off on ImageNet-256 at SNR = 10 dB.}
	\label{fig:Latency Quality}
	\vspace{-4pt}
\end{figure}

To further substantiate the efficacy of the proposed scheduling mechanism, we analyze the distribution of transmission latency and the latency–quality trade-off achieved by CAPS-TDPC. Fig. \ref{fig:Latency CDF} illustrates the cumulative distribution function (CDF) of transmission latency under Rayleigh fading channels. The fixed-timestep baseline exhibits a constant latency of 500 steps, whereas CAPS-TDPC adaptively determines the transmission time based on the scheduling metric. The CDF curve indicates that CAPS-TDPC attains a median latency of approximately 140 steps, representing a 72$\%$ reduction relative to the fixed baseline. This substantial decrease demonstrates that the scheduler effectively leverages favorable channel conditions to transmit intermediate semantic features earlier, thereby circumventing the prolonged waiting time associated with full diffusion completion.

Fig. \ref{fig:Latency Quality} depicts the latency–quality trade-off on the ImageNet dataset at an SNR of 10 dB. The fixed baseline yields a single operating point, whereas the proposed generates method produces wide broad spectrum latency–PSNR pairs, with an mean latency of 159~steps and an average PSNR of 25.43~dB. Compared to fixed baseline, this corresponds to represents latency 68$\%$ reduction at cost of expense a 0.77~dB PSNR degradation. Notably, points distribution that CAPS-TDPC demonstrates flexibly allocate transmission dynamically according to instantaneous based on and semantic conditions when the evolution. Under favorable channel conditions, early transmission can be achieved with only a moderate reduction in quality. Conversely, when channel conditions are suboptimal, the system delays transmission to obtain more refined features, thereby maintaining reconstruction quality at the expense of increased latency. This prior to trade-off enables intrinsic to allows to diverse adapt effectively requirements, varying it particularly rendering for latency-sensitive well-suited in mobile environments.

Collectively, these findings demonstrate that the scheduling mechanism not only substantially decreases the transmission latency but also offers a flexible latency–quality trade-off that can be adapted to diverse channel conditions and application requirements. Furthermore, the patch-wise adaptive allocation enhances this flexibility by prioritizing computational resources toward semantically significant and readily recoverable regions, as examined in the subsequent subsection.

\subsubsection{Ablation Studies on Key Components}

To assess the contribution of each core component within the proposed CAPS-TDPC framework, we performed ablation experiments using the ImageNet dataset under an AWGN channel at an SNR of 10 dB. The results, summarized in Table~\ref{tab:ablation}, compare the complete CAPS-TDPC system with several variants in which individual modules were either disabled or replaced.

The complete CAPS-TDPC framework attains the highest reconstruction quality while sustaining a low average latency of 128 steps, representing a 74.4$\%$ reduction relative to the fixed-timestep baseline. The removal of the scheduler results in a deterioration across all quality metrics and abolishes latency reduction, underscoring the critical importance of joint scheduling in balancing transmission delay and semantic fidelity. Removing the scheduler results in a 500-step latency and significant quality degradation, confirming the necessity of adaptive scheduling. Furthermore, the removal of path deficit compensation, semantic attention, or adaptive reverse steps each leads to a notable decline in image quality while maintaining the same average latency of 128 steps, highlighting their respective contributions to reconstruction fidelity.

Further ablation studies demonstrate that the removal of path deficit compensation, semantic attention, or adaptive reverse steps each results in a significant decline in image quality, while maintaining the same average latency as the complete CAPS-TDPC model. These findings confirm that these components are essential for reconstructing high-fidelity images from prematurely transmitted intermediate features. Furthermore, uniform allocation of reverse steps not only degrades image quality but also increases latency to 300 steps, highlighting the importance of semantic-latency-driven adaptive reconstruction.

In summary, the ablation study confirms that the scheduler, path deficit compensation, semantic attention, and adaptive reverse steps are all essential for attaining the superior performance of CAPS-TDPC.

\begin{table}[htbp]
	\centering
	\caption{Ablation study results on ImageNet-256 at SNR = 10 dB.}
	\label{tab:ablation}
	\small
	\setlength{\tabcolsep}{3.5pt}
	\renewcommand{\arraystretch}{0.95}
	\resizebox{\columnwidth}{!}{
		\begin{tabular}{p{2.8cm} 
				S[table-format=2.2] 
				S[table-format=1.3] 
				S[table-format=1.3] 
				S[table-format=3.0]}
			\toprule
			Configuration & {PSNR $\uparrow$} & {MS-SSIM $\uparrow$} & {LPIPS $\downarrow$} & {Latency (steps)} \\
			\midrule
			Full CAPS-TDPC          & \bfseries 28.00 & \bfseries 0.955 & \bfseries 0.055 & \bfseries 128 \\
			Fixed Timestep & 26.25 & 0.921 & 0.079 & 500 \\
			w/o Path Deficit      & 24.06 & 0.882 & 0.117 & 128 \\
			w/o Sem. Attn         & 23.52 & 0.866 & 0.126 & 128 \\
			w/o Adaptive Steps    & 24.87 & 0.896 & 0.104 & 128 \\
			Uniform Allocation    & 24.23 & 0.887 & 0.113 & 300 \\
			\bottomrule
			\vspace{-2pt}
		\end{tabular}
	}
	\footnotesize
	\raggedright
	$\uparrow$: higher is better; $\downarrow$: lower is better.\\
	\emph{w/o Fixed Timestep}: Timestep $t=500$.\\
	\emph{w/o Sem. Attn}: without Semantic Attention.\\
	\emph{Uniform Allocation}: assigns equal reverse steps to all image patches.
\end{table}

\vspace{-2pt}
\section{Conclusion} 
\label{sec:conclusion}

This paper proposes the CAPS-TDPC framework to address the fundamental temporal mismatch between the iterative latency inherent in DMs and the fading characteristic of wireless channels. By prioritizing the channel clock over generation completion, CAPS-TDPC facilitates the early transmission of intermediate semantic features during favorable channel conditions, thereby significantly reducing the end-to-end latency. Furthermore, a path deficit-based compensation mechanism is introduced to quantify the recovery difficulty and enable accurate reconstruction from interrupted diffusion processes. Moreover, a semantic-latency-driven patch-wise adaptive allocation strategy prioritizes computational resources toward regions exhibiting high semantic value and low decoding cost, thereby achieving an efficient balance between quality and latency. Experimental results obtained from multiple datasets under both AWGN and Rayleigh fading channel conditions indicate that the proposed CAPS-TDPC method consistently surpasses existing baseline approaches in terms of PSNR, MS-SSIM, and LPIPS metrics. Furthermore, CAPS-TDPC achieves significant reductions in latency alongside robust performance improvements, thereby substantiating the efficacy of the introduced scheduling and adaptive compensation mechanisms.

\ifCLASSOPTIONcaptionsoff
\newpage
\fi

\vspace{-4pt}
\bibliographystyle{IEEEtran}
\vspace{-2pt}
\bibliography{refer.bib}

\end{document}